# A Review on Controllability of Multi-Agent Systems Using Switched Networks

Javeria Noor

*Control Science and Engineering, School of Automation, Central South University, Changsha 10533, China*

I.  *Abstract—* Controllability refers to a situation in which a Multi-agent System may be steered from one state to another using specified rules. As a result, there is belief in achieving a given condition by explicit advances. The level of dynamism in the topology and the level of determinism in the environment are two fundamental criteria that determine multi-agent system controllability. The topology of a powerful multi-agent system changes on a regular basis, altering the connections between agents and hence their cooperative effort. This survey focuses on the controllability of MAS in a switching network with a leader that follows the closest neighbor collaboration rule. The leader/pioneer is a single agent that functions as an output to control other agents/members. Because the results of activities are unknown under non-deterministic situations, agents must choose new activities after observing the aftereffects of their prior actions, which causes time delay and limits agent proactivity. Controllability is often achieved in a concentrated manner in the literature, where a certain leader educates supporters how to achieve a specific goal. Controllability has different applications which incorporates managing airplane, vehicle, and robots.

**Index words—** Controllability, switched systems, switched networks, multi-agent systems (MAS), nearest-neighbor rule, switching topology.

II.  **Introduction:**

A MAS refers to a group of independent agents who collaborate in a structured environment. Control engineers are enthusiastic about developing processes for a MAS to achieve a given global control goal via circulating detecting, correspondence, registering, and control. The creation of automatic flying vehicles and appealing micro robots are examples of small MAS. Massive MAS include brilliant frameworks, traffic organizations, sensor organizations, natural frameworks, and informal communities, to name a few. Agreement, synchronization, and arrangement are all common global control objectives. The essential feature of MAS is their autonomy, which demands dispersed tasks. It is this element that permits a MAS to be adaptable in the sense that when network size increments, comparable global control targets can in any case be accomplished without expanding the intricacies for detecting, correspondence, figuring and control. All together to accomplish the necessary worldwide control objective, a correspondence network is fundamental for agents to divide data between their neighbors.

*\*The author is with School of Automation, Department of Control Science and Engineering \* Email: 214618005@csu.edu.cn*

The controllability of a MAS with non-identical interaction capacity and productivity among distinct dynamic agents using adjoining or closest neighbor rules with a fixed topology or switching topology has been the focus of this paper. For the MAS's trading controllability, some basic and compelling suitable conditions have been inferred. By properly choosing one of the agents as the leader, some computationally compelling enough requirements have been gained for the system to be controllable, in any case, for the situation when every sub-system is not controllable. The outcomes in [1] give further knowledge into the impact of the common interaction designs on the collective movement of a MAS, even with time delay.

When considering a system with more than one agent, this is usually the case, a system's dynamic behavior is characterized by its trajectory in time, which is indicated by a vector function x(t) of time t. The system is represented by a vector function x1(t, k) of both time t and agent index=k, where t ∈ [0, ∞] is the time variable in a continuous time setup or sequential time instants in a discrete time setup= t = 0, 1, 2, k = 1, 2, 3.... Furthermore, for the agent k, let $x_2$(t, k) represent the network impact, i.e., the external input originating from the network. Now, the dynamic network behavior can be described in two dimensions, namely, the time dimension and the network dimension, using the functions $x_1$(t, k) and $x_2$(t, k). The model is referred to as a two-dimensional (2D) model in this case. On the one hand, the following equation 1 represents the system dynamics in the time dimension.

$$\delta x_1(t,k) = f(x_1(t,k), x_2(t,k), t, k) \qquad (1)$$

where the operator δ is defined as follows

$$\delta x_1(t,k) := \begin{cases} dx_1(t,k)/dt, & \text{continuous time} \\ x_1(t+1,k) - x_1(t,k), & \text{discrete time} \end{cases}$$

On the other hand, the influence in the network dimension is as follows

$$\Delta x_2(t,k) = g(x_1(\tau(t,k), n(t,k)), t, k) - x_2(t, k-1) \qquad (2)$$

The topology of the correspondence network among the agents has a significant impact on whether a distributed network can reach an agreement using a decentralized control law. The properties of the adjoining capacity n (t, k) in network topologies are the subject of research (2). For fixed topologies, the work n (t, k) is independent of t; for exchanging topologies, the capacity n (t, k) fluctuates with t but only accepts values from a small set; and for time-varying topologies, the capacity n (t, k) is specified more broadly, generally depending on the organization state. Additionally, the unique leader-follower topology and exploration on connectivity's preservation are talked about in this segment[2].

In last few years, distributed control for movement coordination of different unique agents has arisen as a very active research area [3-7]. This is due to advancements in calculations, computations, sensing technologies, and detection innovations, as well as the widespread use of MAS in a variety of applications, including flocking motion of multiple mobile agents control of automated air vehicles, tutoring of submerged vehicles, demeanor arrangement of satellite bunches, and clog control of correspondence networks.[8-10]. The worldwide cooperative conduct of biologic masses, where aggregate movements may emerge from gatherings of specialists through neighborhood and restricted connections among them, is basically charged up in this direction. Some mathematically sound conditions are established for such MAS to be controllable. The results show that a MAS can be controlled even if all its subsystems are uncontrollable by selecting one of the agents as the pioneer and planning the neighbor-collaboration rules rationally using an exchange topology. The fixed topology scenario is analyzed, and new controllability conditions and input formulas for optimal network formation are presented. In addition, the controllability of MAS switching networks in the face of correspondence delay is explored.[11]. Displaying modelling and understanding consensus/agreement problems, formation control, flocking motion, and rendezvous have all been investigated in advance. (see, for instance, [12, 13] and the references in these papers).

Researchers, physicists, and system researchers are increasingly interested in the process that underpins fascinating inter-individual relationships among animal groups. Progressive and libertarian association designs are the two most recognized types. Individuals follow their progenitors/leader in the first type, while they follow their neighbors in the second. We show that pigeon rushes take on a mode that flips between the two previously stated systems using high-goal spatiotemporal information derived from a group of pigeons' free flights. Every pigeon will choose its flight path by following the normal/average of its neighbors when travelling in a smooth direction, yet it will shift to follow its leader when sudden curves or crisscrosses occur. Surprisingly, every pigeon considers its neighbors' normal speed while deciding how fast to fly. This switching network is promising for conceivable modern applications in multi-robot system coordination, automated vehicle arrangement control, and different regions [2].

### III. Literature Review:

The controllability issue was brought forward for very first time interestingly for MAS by Tanner in [14], and get advancement in [12, 15-18]. The issue is on how the interconnected networks can be directed to explicit position by managing the movement of a solitary system that assumes the role of the group leader. This is the thing that the supposed the group can be controllable. This requires the depiction of conditions under which the pioneer/leader can move the devotee/follower into any optimal position or plan/setup [15]. That is, to infer conditions for a group of system inter-connected through closest neighbor collaboration rules, to be controllable by one of them going about as a leader [14]. It is basically a sort of development control issue[19].

The issue is changed to a classical notion of controllability in [14] regarding a proper inter-connection topology and a switching controllability issue in [17, 18] as for a switching topology. One of the elements for the controllability issue concentrated in [14, 17, 18] is that the leader is anticipated to be unit-directional, for example, the leader's neighbors must adhere to the inter-connection closest neighbor regulations, but the leader is unconcerned and can choose any agent. Similarly, the agent is uninterested in standard configuration changes and merely acts as an exterior control signal. The member has no effect on the leader, but the leader and other members have an impact on each other. The fixed topology was used to infer movement controls, which is an important step on the way to a more practical unique/dynamic configuration. For instance, in addition to [8, 12], the feasibility issue of accomplishing a predetermined mathematical development of a gathering of unicycles was explored in [20], In order to ensure the closed system's asymptotic convergence, significant and acceptable graphical requirements for the presence of a local information controller were inferred. Our goal is to look at formation control, which is framed as a controllability issue in which the elements are controlled by the switching network and leaders in this study. The main finding is an algebraic characterization of controllability.[21].

The result has the drawback of providing little insight into the effect of dynamic/switching topologies on controllability. The second result then seeks to compensate for this shortcoming, demonstrating that MAS controllability is dependent on the productively planning of a unique evolution design for the topologies of the corresponding dynamic network. The findings aid in a better understanding of the relationship between development control and the rapid evolution of interconnection networks.

Be that as it may, investigation of some central issues concerning the control of MAS, for example, the controllability of MAS, stays to be a difficult errand. In a recent research, Tanner [14] considered a basic inter-connected framework model that comprises of various mobile agents with one-integrator elements, inter-connected through closest neighbor rules. Fundamental and adequate conditions are acquired for such interconnected frameworks to be controllable by a group member going about as the pioneer, which is thought to have the option to influence its neighbors yet not be impacted by other group members. The outcomes are determined dependent on a decent time-invariant closest neighbor topology. Then again, for the instance of networks with switching topologies, as regularly experienced by and by, it is normally hard to manage the controllability issue because of the intricacy of the topology and need of theoretical tools. Up until this point, there have been not many outcomes accessible in the writing for the controllability of MAS with switching technologies in the discrete time model [22]. Another important element that is rarely discussed is the time delays in switching network systems caused by low rates of correspondence/communication among agents. Furthermore, determining the controllability of a powerful organization with temporal delays is unquestionably not a straightforward task.

## IV. Controllability of Switching Networks

This research focuses on a multi-agent network with an autonomous leader in continuous time (specifically, the leader is unaffected by the members while the members are impacted by the leader straightforwardly or in a roundabout way). The member agents are intended to have different levels of collaboration capability or expertise. Every member agent updates its state based on either the current or delayed data available from its bordering members and the leader, and the leader plays the role of an outside input/contribution to manage the global conduct of the interconnected system. For such a multi-agent system with switching network, fixed topology, and time-delayed interconnection, several simple controllability conditions are deduced, accordingly. The conversation depends on the new work on switched control systems [23] and an early outcome on the controllability of time delay system [24].

The basic findings of continuing research reveal that a MAS can be completely controlled even if each of its subsystems (member agents) is uncontrollable by a leader. Models are provided to demonstrate how the system can grasp the desired final configuration by developing suitable interconnection configurations. The network in the provided setup is consistently not controllable if the leader is following up on each part with a comparable input, independent of the connectivity contents of the members themselves, for the fixed topology situation. As a result, the leader has no influence over a set of agents connected by fixed and identical couplings. This sums up an outcome acquired by Tanner [14] under a total chart condition on the network's topology, which is an extraordinary instance of the network model considered in this paper [14].

The following notations will be used throughout this paper: Z and R denote the set of integer and real number, $R^N$ denotes the $N^{th}$−dimensional real-vector space

Consider a group of N +1 dynamic specialists (with unit mass) travelling in an n-dimensional Euclidean space, where the agent with the index of 0 is designated as the leader and the other agents with the index of 1, N, are referred to as individuals. Individuals have no effect on the leader, but the leader and its neighbors may have an impact on every portion. Equation 3 represents a continuous time kinematic model of agent gathering as

$$\dot{x}_i(t) = -\sum_{j \in \mathcal{N}_i} w_{ij}(x_i(t) - x_j(t)) - \gamma_i w_{i0}(x_i(t) - x_0(t)), i = 1, \dots, N \quad (3)$$

where $x_i \in R^1$ is the state of agent i; W =[$w_{ij}$]∈$R^{N \times N}$ is the coupling matrix with each $w_{ij}$>0 and $W_{ii}$=0; $w_{i0}$>0 are the coupling weights between the leader and the members, and $w_0 = (w_{10}, w_{20} \dots w_{N0})^T \in R^N$ i =1 if 0~i or 0 otherwise. For simplicity in notation, here only the case of n =1 is considered, although the discussion is applicable to the general case of n>1. Let x = ($x_1, \dots, x_N$) T be the stack vector of all the agent states, so it follows that equation (4)

$$\dot{x}(t) = Fx(t) + rx_0(t) \quad (4)$$

Specifically, the idea of 'exchanging controllability' is utilized all through, which is the conventional controllability of new type of system that has a switched topology. A numerically more exact definition is as per the following.

**A non-zero state x of (2) is switching controllable if there exists an integer M<∞ and switching sequence $\lambda = (F_{k_m}, r_{k_m})^M_m$, where $k_m \in K$ is the index of the $m^{th}$ realization $(k_m, h_m)$ with $h_m > 0$ being the time interval of $(F_{k_m}, r_{k_m})$, , and a piecewise continuous input $x_0(t)$, during $[0, T]$, where $T = €^M_m h_m$, such that $x(0) = x$ and $x(T) = 0$. If all non-zero states x in (3) are switching controllable, then system (3) is said to be switching controllable**.

This legitimacy is a lot of attractive in applications as it can give more opportunity to the design of the MAS. This additionally provides help for the designs of a switching path to ensure the switching controllability of the MAS[14].

In last few years, a lot of efforts has been made for developing complicated dynamics control algorithms for MAS including heterogenous, non-linear and uncertain dynamics. These outcomes have been obtained in a relatively simple framework, especially in a fixed and connected topology. Due to various factors like hardware limitation and external interference, topology of communication network is not always fixed. Therefore, consensus of multi-agent systems in a time varying network gets lots of attention [25, 26]. In technical point of view, Consensus of multi-agent systems of switched networks e.g. (network topology and communication weight are switched over a finite set, relies on stabilization development technique for switched systems.

In [25] consensus problem for nonlinear heterogenous and uncertain multi agent systems in switching networks is solved using input to state stability of switched system. Results of consensus is guaranteed under so called jointly connected assumptions which does not need measurements for agents' state or fast switching.

Nonlinear heterogenous multi agent systems 's output synchronization with switched network is also considered in this survey. For nonlinear heterogenous multi agent systems, Output synchronization of switching network of time varying directed graphs is studied in [27]. output synchronization of nonlinear heterogenous multi agent systems with switching networks is not easy in a sense that none of graph is assumed to be connected (means containing spanning tree) Within two step networks for output synchronization of nonlinear heterogenous MAS. The main technical challenge lies in the existence of switching strategies under which a class of switched unstable subsystems subject to external perturbation can be input-to-state stabilized. A controller design approach is proposed in [27] for guarantee of the existence of such switching strategies.

Class of positive switched system's an optimal control problem is also studied in this survey. This paper deals with the optimal control of a class of positive switched systems. The important feature of this class is that only the diagonal entries of the dynamic matrix are altered by switching process. The control input is represented by the switching signal itself and the optimal control problem is that of minimizing a positive linear combination of the final state variable. First, the switched system is embedded in the class of bilinear systems with control variables living in a simplex, for each time point. The main result is that the cost is convex with respect to the control variables. Any Pontryagin solution will be optimal because of this. The best approach is then found using algorithms, and an example from a simple model for HIV mutation mitigation is discussed.[28]. The Pontryagin method is used to perform a useful search for function, reducing the number of computer-based experiments significantly. There is a Pontryagin minimum principle that is used for optimal solution, as well as a Pontryagin maximum principle that can be seen as an extension of the COV and is widely used to obtain the strategy for optimal control of continuous processes.[29-33].

## v. Conclusion:

In this survey of the controllability of switched networks of MAS, we focus on the controllability of MAS in the leader-follower structure, in which the leader/pioneer acts as a control input and the followers are interconnected using nearest/nearest-neighbor rules. Fundamental and additionally adequate conditions are inferred, as well as powerful evolution designs that are usefully intended for the controllable system. The findings suggest that the advancement of dynamically interconnected topologies may have a significant impact on the arrangement of MAS. The consensus problem for nonlinear heterogeneous and uncertain multi agent systems in switching networks using input to state stability of switched systems, as well as Nonlinear heterogeneous multi agent systems' output synchronization, are also discussed in this paper.